\documentclass[nofootinbib,prl,aps,reprint,amsmath,amssymb]{revtex4-1}
\usepackage{hyperref}
\usepackage{changes}
\usepackage[section]{placeins}
\usepackage[titletoc]{appendix}
\usepackage{xcolor}
\usepackage{dsfont}
\usepackage{bm}
\usepackage{mathtools}
\usepackage{enumerate}
\DeclareMathAlphabet{\mathpzc}{OT1}{pzc}{m}{it}

\newcommand{\sayy}[1]{`#1'}
\DeclarePairedDelimiter\abs{\lvert}{\rvert}%
\providecommand{\href}[2]{#2}

\makeatletter
\newcommand*{\rom}[1]{\expandafter\@slowromancap\romannumeral #1@}
\makeatother

\def\be{\begin{equation}}
\def\ee{\end{equation}}
\def\bea{\begin{eqnarray}}
\def\eea{\end{eqnarray}}

\def\sig{\sigma}

\def\la{\langle}
\def\ra{\rangle}
\def\Eu{ \mathfrak{H} }

\def\obs{\mathcal{O}}
\def\emi{\mathcal{E}}

\definecolor{MyB}{rgb}{0.1,0.1,1.0}

\begin{document}
\title{Redshift drift cosmography for model-independent cosmological inference} 
\author{Asta~Heinesen}
\email{asta.heinesen@ens--lyon.fr}
\affiliation{Univ Lyon, Ens de Lyon, Univ Lyon1, CNRS, Centre de Recherche Astrophysique de Lyon UMR5574, F--69007, Lyon, France}

\begin{abstract} 
We develop a cosmographic framework for analysing redshift drift signals of nearby sources model-independently, i.e., without making assumptions about the metric description of the Universe. 
We show that the Friedmann-Lema\^{\i}tre-Robertson-Walker (FLRW) prediction is altered nontrivially by regional anisotropies and inhomogeneities. 
In particular, we find that the position drift of the sources is nontrivially linked to the redshift drift signal. 
The redshift drift signal for nearby sources might be formulated in terms of an effective deceleration parameter, which reduces to the FLRW deceleration parameter in the homogeneous and isotropic limit.  
The presented cosmographic framework can be used for model-independent data analysis, exploiting the fact that the exact anisotropic redshift drift signal at lowest order in redshift is given by a \emph{finite} set of physically interpretable coefficients. 
We discuss physical limits of interest as well as challenges related to the framework. 
\end{abstract}
\keywords{Redshift drift, relativistic cosmology, observational cosmology} 

\maketitle

\section{Introduction}
Cosmological data analysis has historically relied on exact symmetry assumptions in order to infer information about the kinematics of the Universe. 
Isotropic modelling assumptions go back to the founding of general-relativistic cosmology, and to the first measurements indicating the expansion of space \cite{Lemaitre:1927,Slipher:1917,Hubble:1929}. 
The Lambda Cold Dark Matter ($\Lambda$CDM) cosmological paradigm is based on the class of homogeneous and isotropic Friedmann-Lema\^{\i}tre-Robertson-Walker (FLRW) models, 
and the majority of modern cosmological data analyses are carried out assuming a class of FLRW models for describing observables. 
When cosmological data is analysed within the $\Lambda$CDM model, a number of tensions\footnote{The \sayy{Hubble tension} \cite{Riess:2019cxk,DiValentino:2021izs} is perhaps the most significant tension currently, and arises from disagreement between the FLRW Hubble parameter as determined by nearby probes and by data from the cosmic microwave background (CMB).} and \sayy{curiosities} emerge \cite{Buchert:2015wwr,Perivolaropoulos:2021jda}. 
Such tensions, if not due to unaccounted-for astrophysical biases, might be a sign that the precision in cosmological data has surpassed the accuracy of the dynamical space-time model employed. 
With next-generation surveys we will be able to make precise cosmological measurements, which will require an equal precision in theoretical modelling and schemes for data analysis. 

Optical drift effects \cite{Quercellini:2008ty,Liske:2008ph,Quercellini:2010zr,Korzynski:2017nas,Bolejko:2019vni} are temporal changes of cosmological observables such as angular position, redshift, flux, and luminosity distance, which are detected by measuring the same astrophysical sources over time. 
Since lifetimes of experiments on Earth -- typically of the order of a few decades -- are very small as compared to cosmological time scales of gigayears, the detection of optical drift effects requires extreme precision.  %
Due to the great recent improvement in facilities for observation to meet the required precision, measurements of cosmological drift effects are within reach. 
In particular, the redshift drift effect \cite{Sandage,McVittie,Loeb:1998bu} has received attention as a cornerstone of near-future observations.  
Modern instruments, such as the Extremely Large Telescope (ELT) \cite{2005Msngr122,Martins:2019gxw} and the Square Kilometer Array (SKA) \cite{SKA:2018ckk,Maartens:2015mra,Klockner:2015rqa}, are estimated to require one to a few decades of observation time for detection of the redshift drift signal \cite{Balbi:2007fx,Liske:2008ph,Alves:2019hrg}. Forecasts relating to Phase2 of SKA predict observation times down to $\sim$ 0.5 years for significant detections of redshift drift, albeit these estimates are associated with a high level of uncertainty \cite{Alves:2019hrg}. 
Flux drift effects could be detected within a few decades with SKA and ELT \cite{Bolejko:2019vni}, while position drift effects are detectable within the same time frame using data from the Gaia observatory \cite{Quartin:2010,Krasinski:2010rc}. 

The direct detection of changes in cosmological observables with time opens the door for model-independent detection strategies of kinematic properties of our Universe.
However, most existing papers concerning the upcoming optical drift measurements assume the FLRW class of models; however, see, e.g.,  \cite{Uzan:2008qp,Balcerzak:2012bv,Rasanen:2013swa,Fleury:2014rea,Hellaby:2017soj,Korzynski:2017nas,Korzynski:2019oal,Marcori:2018cwn,Koksbang:2020zej,Heinesen:2020pms,Heinesen:2021nrc} for theoretical and numerical investigations of the redshift drift signal in more general settings. 

Cosmography without assumptions on field equations or space-time geometry \cite{KristianSachs,ELLIS1985315,EllisMacCallum,Umeh:2013UCT,Clarkson:2011br,Clarkson:2011uk,Heinesen:2020bej} is a powerful tool for fully model-independent data analysis when the quality of data and sky coverage is sufficient. 
So far, no cosmographic framework for analysing redshift drift data without assumptions about the space-time geometry has been developed.   
In this paper we formulate such a framework for model-independent inference of cosmological kinematics and curvature using redshift drift measurements of nearby astronomical sources and position drift measurements of the same sources. 
Our derivations rely on multipole expansion techniques for the redshift drift signal, which were first considered in \cite{Marcori:2018cwn} for FLRW space-times with noncomoving observers and Bianchi \rom{1} space-times, and later formulated in an arbitrary space-time congruence setting \cite{Heinesen:2020pms,Heinesen:2021nrc}.

\newpage
\noindent
\underbar{Notation and conventions:}
Units are used in which $c=1$. Greek letters $\mu, \nu, \ldots$ label space-time
indices in a general basis. The signature of the space-time metric $g_{\mu \nu}$ is $(- + + +)$ and the connection $\nabla_\mu$ is the Levi-Civita connection. 
Round brackets $(\, )$ containing indices denote symmetrisation in the involved indices and square brackets $[\, ]$ denote anti-symmetrisation. 
Bold notation $\bm V$ for the basis-free representation of vectors $V^\mu$ is used occasionally. 

\section{Redshift drift signal for a general space-time congruence} 
\label{sec:review} 
We first review the formulation of the redshift drift signal in terms of a multipole representation in the general setting, making no assumptions about the metric tensor of space-time or the field theory determining the metric. 
Following \cite{Heinesen:2020pms,Heinesen:2021nrc} we consider a general space-time with an unconstrained congruence of emitters and observers (denoted the \sayy{observer congruence} in the following), associated with the 4-velocity field $\bm u$, a proper time function $\tau$, and with kinematic decomposition 
\bea
\label{def:expu}
&& \nabla_{\nu}u_\mu  = \frac{1}{3}\theta h_{\mu \nu }+\sig_{\mu \nu} + \omega_{\mu \nu} - u_\nu a_\mu  \ , \nonumber \\ 
&& \theta \equiv \nabla_{\mu}u^{\mu} \, ,  \quad \sig_{\mu \nu} \equiv h_{ \la \nu  }^{\, \beta}  h_{  \mu \ra }^{\, \alpha } \nabla_{ \beta }u_{\alpha  }  \, , \nonumber \\ 
&& \omega_{\mu \nu} \equiv h_{  \nu  }^{\, \beta}  h_{  \mu }^{\, \alpha }\nabla_{  [ \beta}u_{\alpha ] }   \, , \quad  a^\mu \equiv \dot{u}^\mu \,  , 
\eea 
where $\dot{} \equiv u^\mu \nabla_\mu$ is the directional derivative along the observer congruence flow lines, where $h_{ \mu }^{\; \nu } \equiv u_{ \mu } u^{\nu } + g_{ \mu }^{\; \nu }$ is the spatial projection tensor relative to the observer congruence, and where $\la \ra$ is the traceless and symmetric part of a spatially projected tensor\footnote{See \cite{Spencer:1970} for a general method of decomposing tensors in three dimensions into isotropic and traceless parts, and Appendix A of \cite{Heinesen:2020bej} for the explicit expressions for the decomposition for symmetric tensors with up to six indices.}.  
Let $\bm k$ denote the generator of a geodesic congruence of null rays (henceforth the \sayy{photon congruence}) passing between a pair of causally connected members of the observer congruence. 
We have that 
\bea
\label{Ee}
E \equiv - u^\mu k_\mu  \, , \qquad  e^\mu \equiv u^\mu - \frac{1}{E} k^\mu \, , 
\eea  
denote the photon energy as measured by a member of the observer congruence and the spatial unit-vector of observation of the null ray as seen by the same observer. We introduce the variables 
\bea
\label{positiondrift}
d^\mu \equiv h^{\mu}_{\; \nu} e^\alpha \nabla_\alpha e^\nu \, , \qquad  \kappa^\mu \equiv  h^{ \mu }_{\; \nu }   \dot{e}^\nu \, , 
\eea  
which we denote the \sayy{acceleration vector} and the \sayy{position drift}. The acceleration vector describes the acceleration of $\bm e$ as projected onto the spatial tangent plane defined by $\bm u$. When $\bm e$ is associated with an axis of local rotational symmetry, $\bm d$ vanishes \cite{Elst:1996}, and the norm of $\bm d$ can thus be thought of as a measure of anisotropy around the spatial axis of propagation of the photons. The position drift\footnote{The position drift is equal to the Fermi-Walker transport of $\bm e$ \cite{Korzynski:2017nas} along $\bm u$: $\delta_{\bm u} e^\mu \equiv \dot{e}^\mu + e^\nu F^\mu_{\; \nu} = \kappa^\mu$, with $F^\mu_{\; \nu} \equiv -u^\mu a_\nu + a^\mu u_\nu$.} describes the shift of spatial direction of incoming light from the emitting source as seen relative to an unrotated coordinate system in the observer's frame; see section 4.2 in \cite{Korzynski:2017nas} for the relation between position drift and classical parallax in a generic perturbative setting. 

The drift of the redshift, $z$, as associated with emitters and observers of the congruence description and measured in proper time of the observer can be represented by the integral \cite{Heinesen:2021nrc}  
\bea
\label{redshiftdriftint}
\frac{d z}{d \tau} \Bigr\rvert_{\obs} = E_\emi \! \! \int_{\lambda_\emi}^{\lambda_\obs} \! \! d \lambda \,   \Pi     \, , \qquad z \equiv \frac{E_\emi}{E_\obs} - 1
\eea  
where $\lambda$ is an affine parameter along the null geodesics of the photon congruence with $k^\mu \nabla_\mu \lambda = 1$, and where subscripts $\emi$ and $\obs$ denote evaluation at the points of emission and observation. The integrand, $\Pi$, is given by the \emph{exact} series expansion in $\bm e$ and $\bm d$ \cite{Heinesen:2021nrc} 
\bea
\label{Pimulti}
\hspace*{-0.65cm} \Pi &=& \Pi^{\it{o}}    +  e^\mu \Pi^{\bm{e}}_\mu  + d^\mu \Pi^{\bm{d}}_\mu   +       e^\mu   e^\nu \Pi^{\bm{ee}}_{\mu \nu} + e^\mu   d^\nu \Pi^{\bm{ed}}_{\mu \nu}     \nonumber  \\    
&& +  \, e^\mu   e^\nu  e^\rho \Pi^{\bm{eee}}_{\mu \nu \rho}   +    e^\mu   e^\nu  e^\rho   e^\kappa \Pi^{\bm{eeee}}_{\mu \nu \rho \kappa} \, 
\eea  
with coefficients 
\bea
\label{Picoef}
&& \Pi^{\it{o}} \equiv  - \frac{1}{3} u^\mu u^\nu R_{\mu \nu} -  d^\mu d_\mu   \nonumber  \\    
&& \qquad \; \;      + \frac{1}{3}D_{\mu} a^{\mu} -  \frac{1}{3} a^\mu a_\mu - \frac{3}{5} \sigma^{\mu \nu} \sigma_{\mu \nu} - \omega^{\mu \nu} \omega_{\mu \nu}    \, , \nonumber  \\   
&&  \Pi^{\bm{e}}_\mu  \equiv   - \frac{1}{3}  \theta a_\mu    +  \frac{7}{5}  a_{ \nu} \sigma^{\nu }_{\, \mu} - a^\nu \omega_{\mu \nu} - h^{\nu}_{\mu} \dot{a}_\nu \, ,  \nonumber  \\   
&& \Pi^{\bm{d}}_\mu \equiv - 2  a_\mu \, ,   \nonumber  \\  
&& \Pi^{\bm{ee}}_{\mu \nu} \equiv   2  a_{\la \mu}a_{\nu \ra }       - \frac{9}{7} \sigma_{\alpha \la \mu} \sigma^\alpha_{\; \nu \ra }  - 3 \omega_{\alpha \la \mu} \omega^\alpha_{\; \nu \ra }   - 6 \sigma_{\alpha  \mu} \omega^\alpha_{\; \nu }   \nonumber \\ 
&& \qquad  \; \;   + D_{  \la \mu} a_{\nu \ra }  -  u^\rho u^\sigma  C_{\rho \mu \sigma \nu}   -  \frac{1}{2} h^{\alpha}_{\, \la \mu} h^{\beta}_{\, \nu \ra}  R_{ \alpha \beta }  \,   , \nonumber  \\   
&& \Pi^{\bm{ed}}_{\mu \nu} \equiv  4 (\sigma_{\mu \nu}  - \omega_{\mu \nu}   )  \, ,  \nonumber  \\   
&& \Pi^{\bm{eee}}_{\mu \nu \rho} \equiv  - 4  a_{ \la \mu} \sigma_{\nu \rho \ra }  \, , \nonumber  \\   
&& \Pi^{\bm{eeee}}_{\mu \nu \rho \kappa} \equiv 3 \sigma_{\la \mu \nu } \sigma_{\rho \kappa \ra}  \, , 
\eea  
where $R_{\mu \nu}$ is the Ricci curvature tensor, $C_{\rho \mu \sigma \nu}$ is the Weyl curvature tensor,  the operator $D_\mu$ is the spatial covariant derivative\footnote{The acting of $D_\mu$ on a tensor field $T_{\nu_1 , .. , \nu_n }^{\qquad  \gamma_1 , .. , \gamma_m }$ is defined as: $D_{\mu} T_{\nu_1 ,  .. , \nu_n }^{\qquad  \gamma_1 , .. , \gamma_m } \equiv  h_{ \nu_1 }^{\, \alpha_1 } .. h_{ \nu_n }^{\, \alpha_n }    \,  h_{ \beta_1 }^{\, \gamma_1 } .. h_{ \beta_m }^{\, \gamma_m }    \, h_{ \mu }^{\, \sigma } \nabla_\sigma  T_{\alpha_1 ,  .. , \alpha_n }^{\qquad \beta_1 , .. , \beta_m }$ .
}, and where the decomposition of the terms in the series expansion has been made such that coefficients with more than one space-time index are traceless. 

\section{Cosmography for model-independent data analysis of nearby sources} 
\label{sec:modelindependent}
The representation (\ref{Pimulti}) of $\Pi$ in terms of a multipole expansion in the vectors $(\bm e,\bm d)$ is not directly suitable for formulating a model-independent redshift drift cosmography for observational analysis, since the acceleration vector $\bm d_\obs$ is in general not a measurable quantity. 
We make use of the relation \cite{Heinesen:2020bej} 
\bea
\label{kderive}
\hspace*{-0.5cm} \frac{k^\nu \nabla_\nu e^\mu}{E} &=&  (e^\mu \! - \! u^\mu) \Eu   -   e^\nu \! \! \left(\! \frac{1}{3} \theta h^{\mu}_{\; \nu}  + \sigma^{ \mu}_{\; \nu} + \omega^{\mu}_{\; \nu} \! \right) +  a^\mu \,  
\eea  
and the definition of $\bm e$ given in (\ref{Ee}) to rewrite $\bm d$ in terms of $\bm e$ and $\bm \kappa$ in the following way:  
\bea
\label{drewrite}
\hspace*{-0.4cm} d^\mu  =  \kappa^\mu - a^\mu + e^\nu (\sigma^\mu_{\; \nu}  + \omega^\mu_{\; \nu} )    + e^\mu e^\nu a_\nu - e^\mu e^\nu e^\rho \sigma_{\nu \rho}  \, .  
\eea  
{The realisation that $\bm d$ can be rewritten in terms of $\bm \kappa$, $\bm e$, and kinematic quantities of the observer congruence is crucial for the formulation of a general cosmographic expression that is applicable for data analysis.   
This is because the position drift vector as evaluated at the observer location $\bm \kappa_\obs$, unlike the acceleration vector $\bm d_\obs$, is a measurable quantity (see the discussion below), which in turn makes it a good expansion variable for the anisotropic cosmography. }

Substituting $\bm d$ with (\ref{drewrite}) in the series expansion (\ref{Pimulti}) and rearranging terms, $\Pi$ can be written as the following series expansion in $\bm e$ and $\bm \kappa$: 
\bea
\label{Pimultikappa}
\hspace*{-0.65cm} \Pi &=&  -  \kappa^\mu \kappa_\mu  + \Sigma^{\it{o}}    +  e^\mu \Sigma^{\bm{e}}_\mu    +       e^\mu   e^\nu \Sigma^{\bm{ee}}_{\mu \nu} + e^\mu   \kappa^\nu \Sigma^{\bm{e\kappa}}_{\mu \nu}    
\eea  
with coefficients 
\bea
\label{Picoefkappa}
&& \Sigma^{\it{o}} \equiv  - \frac{1}{3} u^\mu u^\nu R_{\mu \nu}     + \frac{1}{3}D_{\mu} a^{\mu} + \frac{1}{3} a^\mu a_\mu    \, , \nonumber  \\   
&&  \Sigma^{\bm{e}}_\mu  \equiv   - \frac{1}{3}  \theta a_\mu    -   a^{ \nu} \sigma_{\mu \nu}  + 3 a^{ \nu} \omega_{\mu \nu}    - h^{\nu}_{\mu} \dot{a}_\nu \, ,  \nonumber  \\   
&& \Sigma^{\bm{ee}}_{\mu \nu} \equiv     a_{\la \mu}a_{\nu \ra }  + D_{  \la \mu} a_{\nu \ra }    -  u^\rho u^\sigma  C_{\rho \mu \sigma \nu}   -  \frac{1}{2} h^{\alpha}_{\, \la \mu} h^{\beta}_{\, \nu \ra}  R_{ \alpha \beta }  \,   , \nonumber  \\   
&& \Sigma^{\bm{e\kappa}}_{\mu \nu} \equiv  2 (\sigma_{\mu \nu}  - \omega_{\mu \nu}   )  \, . 
\eea  
The representation (\ref{Pimultikappa}) of $\Pi$ in terms of the position drift $\bm \kappa$ is a simple truncated expression at quadratic order in $(\bm e, \bm \kappa)$. (Compare to the representation (\ref{Pimulti}) with expansion in $(\bm e, \bm d)$, which has terms up to fourth order in $\bm e$.)
This representation is furthermore of observational interest, since position drifts of galaxies are detectable with facilities such as Gaia\footnote{The identification of an irrotational coordinate system is needed for isolating the position drift effect from local kinematics of the observer \cite{Krasinski:2010rc}. Such a non-rotating reference frame might be defined relative to a \sayy{background} of extragalactic sources \cite{Mignard2018GaiaDR}.} \cite{Krasinski:2010rc}. 
It is interesting to note that a general relation between redshift drift and position drift has also been obtained in\footnote{The formalism presented in \cite{Korzynski:2017nas} does not rely on a congruence description of the observers and emitters of light, but instead introduces vector fields from parallel transport of the observer and emitter 4-velocities along the null ray, with respect to which observables are decomposed. Therefore the results found for redshift drift in \cite{Korzynski:2017nas} are not directly comparable to those of the present paper.} \cite{Korzynski:2017nas}.

We now consider the redshift drift signal of nearby sources by expanding the signal in the affine parameter along the incoming null geodesic. 
Expanding (\ref{redshiftdriftint}) around the point of observation $\obs$, the redshift drift signal to lowest order in affine distance along the null ray reads 
\bea
\label{zdfirst}
\frac{d z}{d \tau} \Bigr\rvert_{\obs} = - E_\emi \Pi_\obs \Delta \lambda  + \mathcal{O}( \Delta \lambda^2)   \, , 
\eea  
where $\Delta \lambda \equiv \lambda_\emi - \lambda_\obs$. 
We can rewrite (\ref{zdfirst}) as a first-order Taylor series expansion in redshift by noting that 
\bea
\label{def:Eevolution}
\frac{ {\rm d} E}{{\rm d} \lambda} = - k^{\mu}\nabla_{\mu} ( k^{\nu} u_\nu ) = - E^2  \Eu \, ,  
\eea 
with 
\bea
\label{def:h}
\Eu \equiv  \frac{1}{3}\theta  - e^\mu a_\mu + e^\mu e^\nu \sigma_{\mu \nu}  \, , 
\eea 
which, assuming that the function $\lambda \mapsto z(\lambda)$ is invertible, gives \cite{Heinesen:2020bej}
\bea
\label{lambdacoef}
\Delta \lambda =   - \frac{1}{E_\obs \Eu_\obs} z + \mathcal{O}(z^2) \, 
\eea 
along each null ray. 
Using (\ref{lambdacoef}) in (\ref{zdfirst}), we finally have the first-order cosmographic expression for redshift drift 
\bea
\label{zdfirst2}
\frac{d z}{d \tau} \Bigr\rvert_{\obs} =  - \mathbb{Q}_\obs \Eu_\obs z + \mathcal{O}(z^2) \, , \qquad \mathbb{Q} \equiv - \Pi/\Eu^2  \, .  
\eea  
Thus, we see that the redshift drift signal of nearby sources is given by $\Eu_\obs$ and $\Pi_\obs$, which are \emph{truncated} series expansions in the direction of incoming light from the source $\bm e_\obs$ and its position drift $\bm \kappa_\obs$. We note that this truncation is \emph{exact}, and that no model assumptions have been used in the expressions (\ref{Pimultikappa}), (\ref{Picoefkappa}) and (\ref{def:h}). 

We might identify $\Eu$ as a natural generalised \sayy{Hubble parameter} prescribing the evolution of photon energy along null rays\footnote{The indentification of $\Eu$ as a generalisation of the FLRW Hubble parameter can further be motivated by noting that the first-order term in the taylor series expansion of luminosity distance is given by $z/\Eu $, as detailed in \cite{Clarkson:2011uk}.}, and identify $\mathbb{Q}$ as an effective observational deceleration parameter for the redshift drift signal as motivated by the well-known FLRW limit of (\ref{zdfirst2}): $- q_{0} H_0 z + \mathcal{O}(z^2)$, where $q_0$ and $H_0$ are the present-epoch deceleration parameter and Hubble parameter of the FLRW model \cite{Neben2012BEYONDH0,Lobo:2020hcz}. We note that an alternative effective observational deceleration parameter, $\mathfrak{Q}$, has been defined from the expansion of luminosity distance $d_L = z/\Eu_\obs + (1- \mathfrak{Q}_\obs) z^2 /(2\Eu_\obs)  + \mathcal{O}(z^3)$ valid for general space-time geometries \cite{Umeh:2013UCT,Clarkson:2011uk,Heinesen:2020bej}. 
Thus, the redshift drift data and distance-redshift data motivate two different generalisations of the FLRW deceleration parameter. This is not surprising, since we might expect different observations to be sensitive to inhomogeneities in the underlying space-time solution in different ways. 

Equation (\ref{zdfirst2}), together with (\ref{Pimultikappa}), (\ref{Picoefkappa}) and (\ref{def:h}), is the \emph{main result of this paper}. 
In the $\mathcal{O}(z)$ vicinity of the observer, the general expression for redshift drift is given in terms of a \emph{finite} number of physically interpretable coefficients, which can be measured given sufficient data and sky coverage: 
given data $\bm e_\obs$ and $\bm \kappa_\obs$ for each source, the effective observational Hubble parameter $\Eu_\obs$ represents nine degrees of freedom\footnote{Spatial vectors represent three degrees of freedom, whereas symmetric and trace-free spatial tensors with two indices represent five degrees of freedom.}, $\{\theta, a_\mu , \sigma_{\mu \nu}\} \rvert_{\obs}$, while $\Pi_\obs$ represents 12 independent degrees of freedom\footnote{The shear term in $\Sigma^{\bm{e\kappa}}_{\mu \nu} \rvert_{\obs}$ is specified by the quadrupole moment $\sigma_{\mu \nu} \rvert_{\obs}$ of $\Eu_\obs$. This leaves $\omega_{\mu \nu} \rvert_{\obs}$ to be determined independently, which is specified by the three degrees of freedom of a three-dimensional antisymmetric tensor with two indices.} 
, $\{\Sigma^{\it{o}}  , \Sigma^{\bm{e}}_\mu  , \Sigma^{\bm{ee}}_{\mu \nu}, \Sigma^{\bm{e\kappa}}_{\mu \nu}\} \rvert_{\obs}$. 

In a fully model-independent analysis, the coefficients $\{\theta, a_\mu , \sigma_{\mu \nu}\} \rvert_{\obs}$ and $\{\Sigma^{\it{o}} , \Sigma^{\bm{e}}_\mu  , \Sigma^{\bm{ee}}_{\mu \nu}, \Sigma^{\bm{e\kappa}}_{\mu \nu}\} \rvert_{\obs}$ are to be treated as free parameters giving a total of 21 independent degrees of freedom to be determined from data. 
However, complementary data\footnote{If for instance the coefficients $\{\theta, a_\mu , \sigma_{\mu \nu}\} \rvert_{\obs}$ of $\Eu_\obs$ have already been measured by analysis of distance-redshift data as proposed in \cite{Heinesen:2020bej}, this leaves the coefficients $\{\Sigma^{\it{o}}  , \Sigma^{\bm{e}}_\mu  , \Sigma^{\bm{ee}}_{\mu \nu}, \Sigma^{\bm{e\kappa}}_{\mu \nu}\} \rvert_{\obs}$ to be determined.} or physically motivated assumptions can reduce the number of independent parameters. 
In the following section we shall consider approximations that reduce the number of degrees of freedom in various situations.   

\section{Physically motivated approximations} 
\label{sec:limits}
We shall now discuss limits of the general cosmographic expression for redshift drift (\ref{zdfirst2}). We first discuss the \sayy{monopole limit} relevant for when the sources in the cosmological survey are sufficiently uniformly distributed over the observer's sky.  
Next we shall discuss the potential of measuring the cosmological position drift effect with upcoming surveys, and perform an order-of-magnitude analysis of the terms in the expansion (\ref{Pimultikappa}) involving the position drift. 
We then discuss the limiting case $\bm \kappa_\obs = \bm 0$ and $\bm a_\obs =  \bm 0$, valid for when position drift is subdominant in the expression for redshift drift and the observer congruence is well approximated as being geodesic. 

\subsection{Monopole limit}
We might analyse the monopole limit $\Eu_\obs \rightarrow \theta_\obs/3$ and $\Pi_\obs \rightarrow  -  \kappa^\mu \kappa_\mu \rvert_{\obs}  + \Sigma_\obs^{\it{o}}$, corresponding to the situation where the redshift drift signal is isotropic (independent of $\bm e_\obs$) as seen by the observer $\frac{d z}{d \tau} \rvert_{\obs} \rightarrow  3 \frac{-\kappa^\mu \kappa_\mu \rvert_{\obs}   + \Sigma_\obs^{\it{o}}}{\theta_\obs} z + \mathcal{O}(z^2)$. 
{For a uniform sky distribution of sources in the observer's catalogue, the average redshift drift signal is expected to be well probed by this monopole limit} -- even if higher-order multipoles are significant in the case of individual sources -- since any traceless component is expected to cancel when averaged uniformly over directions.  

{A uniform sky distribution of sources will in practice not be fully obtained in realistic surveys, which are subject to limitations of instrumentation, astrophysical foregrounds, and the underlying anisotropic distribution of galaxies over the observer's sky. 
However, it might in some situations apply as a good lowest-order approximation. 
The limit of a fairly sampled sky is also of theoretical interest, as it constitutes an observationally motivated monopole limit which might be compared with other monopole limits in cosmological modelling, such as that of the FLRW metrics. 
}

The monopole limit of the redshift drift signal leaves the two independent degrees of freedom $\theta_\obs$ and $\Sigma_\obs^{\it{o}}$ to be determined from data\footnote{Under the approximation $\bm \kappa_\obs \! =\! \bm 0$ for each source (see below), the $\mathcal{O}(z)$ monopole signal reduces to $3 \Sigma_\obs^{\it{o}} z/ \theta_\obs $, which renders $\theta_\obs$ and $\Sigma_\obs^{\it{o}}$ degenerate, leaving one effective parameter $\Sigma_\obs^{\it{o}}/ \theta_\obs $ to be determined from data.}. 
In this limit, the redshift drift signal contains contributions from space-time structure via the position drift (see the below discussion on position drift), 4-acceleration of the observer congruence, and the Ricci curvature term $R_{\mu \nu} u^\mu u^\nu$, as can be seen from the expression for $\Sigma_\obs^{\it{o}}$ in (\ref{Picoefkappa}). 
The extrapolation of the FLRW result for redshift drift to the general case is thus nontrivial, even in the monopole limit {of a fairly sampled sky}. 
In particular we note that the isotropised $\mathcal{O}(z)$ redshift drift signal does not in general probe the deceleration of length scales in the observer frame directly. 

The contributions to the monopolar redshift drift signal from position drift are non-positive -- as can be seen directly from (\ref{Pimultikappa}) and (\ref{Picoefkappa}) -- while the 4-acceleration can contribute with terms of either sign, depending on its spatial gradient $D_\mu a^\mu \rvert_{\obs}$. If the 4-acceleration and its spatial gradient are subdominant\footnote{Vanishing 4-acceleration occurs for instance when the energy content of the Universe is well described by a dust source, which is typically assumed to be the case in general-relativistic modelling of the late universe.}, the only way to obtain a positive monopolar redshift drift signal in the $\mathcal{O}(z)$ vicinity of the observer, is thus to have $R_{\mu \nu} u^\mu u^\nu < 0$ which in a general-relativistic context is equivalent to the violation of the strong energy condition. 
This conclusion for low redshift measurements is similar to that found in \cite{Heinesen:2021nrc} for redshift drift signals of sources located at distances much greater than an approximate homogeneity scale, where positive values of redshift drift were found to be likely caused only by violation of the strong energy condition or by a special 4-acceleration profile of the observer congruence.

\subsection{Accounting for position drift} 
We note that a full model-independent analysis requires knowledge of $\bm e_\obs$ and $\bm \kappa_\obs$ for each astrophysical source. 
While the position of the source on the sky, $\bm e_\obs$, is immediately known, its drift, $\bm \kappa_\obs$, is a measurement that is at least as delicate as the redshift drift signal itself. 
Even though position drift effects associated with {light propagation through} large-scale cosmic structure could be determined with upcoming Gaia data \cite{Quartin:2010,Krasinski:2010rc}, it might be nontrivial to combine such detections with detections of redshift drift for a combined analysis of the two effects. 
In the following we shall estimate the magnitude of the position drift effect relative to the expected redshift drift effect from existing order-of-magnitude estimates in the literature. 

\vspace{5pt} 
{\underbar{Model universes with extreme structures:}} \\ 
{We shall now use estimates from extreme universe structures as modelled by LTB and Bianchi metrics as crude upper bounds on position drift. These estimates are complimentary to low-redshift estimates within the perturbed FLRW framework given below.

Cosmological position drifts of galaxies with $z\sim 1$ have been estimated to be of order $\lesssim10^{-6}$ arcsec/year $\approx 10^{-12}$ rad/year for off-center observers} {situated in $\sim 1$ Gpc scale voids as modelled by LTB solutions} \cite{Quartin:2010,Krasinski:2010rc}. 
{Estimates of cosmic position drift within Bianchi I metrics incorporating anisotropic expansion of space with a shearing rate of order 1\% have been found to be $\sim10^{-7}$ arcsec/year $\approx 10^{-13}$ rad/year for sources located at $z\sim 1$ \cite{2009PhRvD..80f3527Q}. 
We might use such estimates as crude upper bounds\footnote{{Since the estimates are provided at $z \sim 1$, we expect the first-order term in the redshift drift cosmography to be a poor approximation of the underlying redshift drift signal on such scales, irrespective of the exact nature of the underlying space-time. The estimates are however still valid for investigating the relative size of various contributions in the first-order term of the series expansion.} } on position drift effects of sources located at $z\sim 1$ in realistic universe models with more moderate structure, and write $\abs{\bm \kappa}_{\obs} \lesssim 10^{-12} \, \text{year}^{-1}$. 

} 

For local expansion rates comparable to current estimates of the \sayy{background} FLRW Hubble parameter: $\Eu_\obs \sim H_0 \sim 10^{-10} \, \text{year}^{-1}$, we have that the contribution from the first term of (\ref{Pimultikappa}) in (\ref{zdfirst2}) is of size $\kappa^\mu \kappa_\mu /  \Eu  \lesssim 10^{-14} \, \text{year}^{-1} \sim 10^{-4} H_0$, {where the evaluation at the point of observation $\obs$ is implicit here and below for ease of notation. }    
Similarly, the coupled term $e^\mu \kappa^\nu \Sigma^{\bm{e\kappa}}_{\mu \nu}$ of (\ref{Pimultikappa}) contributes in (\ref{zdfirst2}) with a term of size $\abs{e^\mu \kappa^\nu \Sigma^{\bm{e\kappa}}_{\mu \nu} / \Eu  }\lesssim 10^{-2} \, \abs{e^\mu \hat{\kappa}^\nu \Sigma^{\bm{e\kappa}}_{\mu \nu}  }=  10^{-2}  \abs{e^\mu \hat{\kappa}^\nu 2 (\sigma_{\mu \nu}  - \omega_{\mu \nu}   )}$, where $\hat{\bm \kappa} \equiv \bm \kappa  / \abs{\bm \kappa}$ is the unit vector aligned with $\bm \kappa$.  

{Assuming that shear and vorticity are subdominant to the isotropised expansion rate, such that\footnote{{This is indeed a conservative estimate in the Bianchi I space-times considered in \cite{2009PhRvD..80f3527Q}, where the shearing rate is $\sim 1\%$.}} $2 \abs{e^\mu \hat{\kappa}^\nu  (\sigma_{\mu \nu}  - \omega_{\mu \nu}   ) } \lesssim 0.1 \Eu $, we have that $\abs{e^\mu \kappa^\nu \Sigma^{\bm{e\kappa}}_{\mu \nu} / \Eu }\lesssim 10^{-3} \, H_0 $. } 
{For comparison, the lowest-order redshift drift signal in the FLRW model is $- q_0 H_0 z$. 
Position drift contributes well below $1 \%$ for $q_0 \approx \! -0.5$ to the $\mathcal{O}(z)$ redshift drift term within these order-of-magnitude estimates when approaching redshifts of unity.    }

\vspace{5pt} 
{\underbar{Perturbed FLRW models in the small redshift regime:}} \\ 
{In first-order FLRW perturbation theory, assuming a dust universe model, cosmological position drift is of order the classical parallax effect \cite{Rasanen:2013swa} $\abs{\bm \kappa_{}} \sim \abs{\bm v} D^{-1}_P $, where $\bm v$ is the spatial velocity of the observer relative to the Poisson frame, and $D_P$ is the parallax distance in the background FLRW model. 
In the low redshift regime, the term proportional to $D^{-1}_P$ dominates, and $D_P  \approx z/H_0$. In this regime, we thus have $\abs{\bm \kappa} \sim \abs{\bm v} H_0 / z$. 

This estimate leads to $\kappa_{}^\mu \kappa_{ \mu} /  \Eu \approx \frac{\abs{\bm v}^2}{z^2}  H_0$, where we have used that $\Eu  \approx H_0$ at lowest order in the linearised perturbative scheme. 
We also have $\abs{e^\mu \kappa_{}^\nu \Sigma^{\bm{e\kappa}}_{\mu \nu} / \Eu } \approx 2 \frac{\abs{\bm v}}{z}  \, \abs{ e^\mu \hat{\kappa}_{}^\nu \nabla_{\mu} v_{\nu} }$, where we have used that the shear and vorticity in the matter frame are given by $\sigma_{\nu \mu}  \approx  \nabla_{\la \mu} v_{\nu \ra} $ and $\omega_{\nu \mu} \approx \nabla_{[\mu} v_{\nu]}$ respectively \cite{vanElst:1998kb} to lowest order in $\bm v$. 
Comparing these coefficients involving position drift to the magnitude of the leading order redshift drift coefficient in the FLRW expression: $\abs{q_0} H_0$, we see that the position drift terms contribute with correction terms of order $2\frac{\abs{\bm v}^2}{z^2}$ and $4 \frac{\abs{\bm v}}{z} \frac{\abs{ e^\mu \hat{\kappa}_{}^\nu \nabla_{\mu} v_{\nu} }}{H_0}$ for $q_0 \approx \! -0.5$. 

For typical cosmological bulk flow velocities $\abs{\bm v} \sim 10^{-3}$ -- representative of for instance the bulk motion of our cosmic neighbourhood at the $\sim 100\,$Mpc scale relative to the CMB frame \cite{2010MNRAS.407.2328F} -- we can thus expect $\sim 1\%$ correction terms on the $\sim 100\,$Mpc scale\footnote{{We use the order-of-magnitude estimate $\abs{ e^\mu \hat{\kappa}_{}^\nu \nabla_{\mu} v_{\nu} }/H_0 \sim \abs{\bm v} / H_0 / (100 \text{Mpc}) \sim 100 \abs{\bm v} $ for the components of the spatial gradient of $\bm v$ on the 100 Mpc scale.}} (corresponding to $z\sim 0.02$).  
{For comparison, a similar order-of-magnitude estimate for the electric Weyl tensor contribution in the quadrupole coefficient (\ref{Picoefkappa}) term gives correction terms of up to order\footnote{{This order-of-magnitude estimate is based on $\abs{u^\rho u^\sigma  C_{\rho \mu \sigma \nu} } \sim \abs{\theta \sigma_{\mu \nu}}$ (see, e.g., equation (111) of \cite{vanElst:1998kb}), and $\abs{\sigma_{\nu \mu}} \approx 
\abs{ \nabla_{\la \mu} v_{\nu \ra}} \sim \abs{ \bm v}/(100 \text{Mpc})$ at the 100 Mpc scale.}} $\frac{\abs{e^\mu e^\nu u^\rho u^\sigma  C_{\rho \mu \sigma \nu}} }{H^2_0} \sim \frac{\abs{e^\mu e^\nu \nabla_{\la \mu} v_{\nu \ra}} }{H_0} \sim 0.1$ at the $\sim 100\,$Mpc scale.  }

}

\subsection{Geodesic and position drift-free approximation} 
Let us suppose that the above relative order of estimates concerning position drift apply to the low-redshift regime of interest, such that we can set $\bm \kappa_\obs = \bm 0$ for a lowest-order description of redshift drift. 
Let us further suppose that the cosmological congruence description is well approximated as being geodesic, such that $\bm a_\obs =  \bm 0$. This is a reasonable assumption above scales of virialized structure, where effective hydrodynamic pressure from velocity dispersion tends to be negligible \cite{Buchert:2005xj}. 

Under this approximation, the expression (\ref{Pimultikappa}) as evaluated at the observer reduces to 
\bea
\label{Pimultikappaapprox}
\hspace*{-0.5cm} \Pi_\obs &=&    \Sigma^{\it{o}}    \rvert_{\obs}    +       e^\mu   e^\nu \Sigma^{\bm{ee}}_{\mu \nu}   \rvert_{\obs}   \quad \text{(limit: $a_\obs^\mu \! = \! 0 $, $\kappa_\obs^\mu \!=\! 0$)}
\eea  
with coefficients 
\bea
\label{Picoefkappaapprox}
&& \Sigma^{\it{o}}  \rvert_{\obs} \equiv  - \frac{1}{3} u^\mu u^\nu R_{\mu \nu}  \rvert_{\obs}     \qquad \text{(limit: $a_\obs^\mu \! = \! 0 $, $\kappa_\obs^\mu \!=\! 0$)}   \, , \nonumber  \\   
&& \Sigma^{\bm{ee}}_{\mu \nu}  \rvert_{\obs}  \equiv     -  u^\rho u^\sigma  C_{\rho \mu \sigma \nu}  \rvert_{\obs}   -  \frac{1}{2} h^{\alpha}_{\, \la \mu} h^{\beta}_{\, \nu \ra}  R_{ \alpha \beta }  \rvert_{\obs} \,   ,  
\eea  
while $\Eu_\obs$ in the case of zero 4-acceleration reads 
\bea
\label{def:happrox}
\Eu_\obs \equiv  \frac{1}{3}\theta_\obs  + e^\mu e^\nu \sigma_{\mu \nu}  \rvert_{\obs}   \qquad \text{(limit: $a_\obs^\mu \! = \! 0 $)}   \, . 
\eea 
In this case we are left with the coefficients $\{\Sigma^{\it{o}} , \Sigma^{\bm{ee}}_{\mu \nu} \} \rvert_{\obs}$ representing six degrees of freedom in addition to the six degrees of freedom incorporated in the coefficients $\{\theta, \sigma_{\mu \nu}\} \rvert_{\obs}$ of $\Eu_\obs$. 
This gives a total of 12 parameters, of which one degree of freedom is redundant\footnote{An overall scaling of the coefficients $\{\Sigma^{\it{o}}  , \Sigma^{\bm{ee}}_{\mu \nu} \} \rvert_{\obs}$ by a scalar can be absorbed into an overall rescaling of $\{\theta, \sigma_{\mu \nu}\} \rvert_{\obs}$ by the inverse of the same scalar, in the case where $\bm \kappa_\obs = \bm 0$.}.  
When the coefficients of $\Eu_\obs$ can be determined by complementary data, for instance by analysis of standardisable candles \cite{Heinesen:2020bej}, the redshift drift signal is given by the six degrees of freedom $\{\Sigma^{\it{o}} , \Sigma^{\bm{ee}}_{\mu \nu} \} \rvert_{\obs}$. 

Let us consider the significance of the coefficients in (\ref{Picoefkappaapprox}). The monopole $\Sigma^{\it{o}}$ is given by the source term, $u^\mu u^\nu R_{\mu \nu}$, of the focusing of the observer congruence worldlines.  
The quadrupole $\Sigma^{\bm{ee}}_{\mu \nu}$ is given in terms of the electric part of the Weyl tensor in the observer frame, $u^\rho u^\sigma  C_{\rho \mu \sigma \nu}$, together with the trace-free part of the spatially projected Ricci tensor, $h^{\alpha}_{\, \la \mu} h^{\beta}_{\, \nu \ra}  R_{ \alpha \beta }$, which are source terms of the evolution of shear in the observer frame. In a general-relativistic hydrodynamical description, the latter term is identified as the anisotropic stress in the observer frame. 
The electric part of the Weyl tensor, $u^\rho u^\sigma  C_{\rho \mu \sigma \nu}$ is not necessarily small in the \sayy{weak field} or \sayy{small relative velocity} limit of gravity. See equation~(113) of \cite{Bruni1992} for the expression for the electric part of the Weyl tensor in linearised perturbation theory around a flat FLRW model. In the Newtonian limit, the electric part of the Weyl tensor translates into the tidal tensor \cite{Ehlers:2009uv}.  

We can further remark that the eigenbases of $\sigma_{\mu \nu}$ and $u^\rho u^\sigma  C_{\rho \mu \sigma \nu}$ coincide in irrotational dust space-times with vanishing divergence of the magnetic Weyl tensor \cite{Maartens:1996uv}. 
Under this model assumption, which might be thought of as a stable approximation in the linear regime of density contrasts \cite{Maartens:1996uv}, the number of independent degrees of freedom represented by $\{\Sigma^{\it{o}} , \Sigma^{\bm{ee}}_{\mu \nu} \} \rvert_{\obs}$ decreases from six to three degrees of freedom.   
Under the present approximation, and independent determination of $\Eu_\obs$, the redshift drift signal at low redshifts directly measures these curvature invariants.

\section{Discussion of upcoming surveys and limitations of the formalism} 
An upcoming probe for the detection of redshift drift is the Lyman-$\alpha$ forest, which is the observed absorption lines in the spectra of quasars coming from the Lyman-$\alpha$ transition of the hydrogen atom \cite{Rauch:1998xn,Balbi:2007fx}. 
The Lyman-$\alpha$ forest is most efficiently probed in the range $2 \lesssim z \lesssim 5$ \cite{Balbi:2007fx,Liske:2008ph}, and is therefore not suitable for the cosmographic framework developed in this paper. 
Another promising experiment is to measure redshift drift with the neutral hydrogen 21-cm emission lines of galaxies, as is the aim with the SKA experiment \cite{Klockner:2015rqa,Martins:2016bbi}. 
Significant detections of the redshift drift signal at $z\lesssim 0.3$ are currently predicted to be within reach in $\sim 40$ years of observation time for SKA Phase1, while for SKA Phase2 the observation time needed for a significant detection could drop to $\sim 0.5$ years \cite{Martins:2016bbi,Alves:2019hrg}. 
See also \cite{Kim:2014uha} for an outline of potential complementary measurements of redshift drift in the low-redshift regime. 

In this paper, we use a congruence description for the emitters and observers of light in the space-time, in order to make a cosmographic representation of redshift drift signals possible. 
The cosmography has the advantage that it can be directly applied to model-independent analysis of redshift drift, as detailed in the above analysis. 

However, the congruence description is also a limitation of the presented framework, as it necessarily implies the (implicit) coarse-graining over scales where caustics in the matter distribution appear.    
Furthermore, regularity of the cosmographic expression (\ref{zdfirst2}) requires invertibility of the function $\lambda \mapsto z(\lambda)$ which prevents the inclusion of physics below and around the scales of virialized structure; see the discussion in \cite{Heinesen:2020bej} in context of luminosity distance cosmography.  
Changes in the redshift signal due to local motion effects {within gravitationally bound structure}s are thus unaccounted for in the presented framework. 

The special-relativistic acceleration of the Solar System relative to an idealised \sayy{background} of extragalactic sources, has been inferred through the secular aberration drift of radio sources and quasars \cite{VLBI,Collaboration2021GaiaED}, with an inferred acceleration smaller than, but of the same order as, the Hubble constant, and might thus be expected to contribute significantly to the redshift drift signal.   

The implications for redshift drift of the Newtonian 3-acceleration of the Solar System relative to a hypothesised frame of idealised Hubble flow was formulated in \cite{Liske:2008ph}, and investigated in detail in \cite{Inoue:2019qvy}, where a dipolar signature in the redshift drift signal with amplitude comparable to the predicted monopole signal from the rate of change of cosmic expansion in the $\Lambda$CDM model was found. 
This predicted dipolar signal is independent of redshift of the sources \cite{Inoue:2019qvy}, and the effect might thus be distinguished from the cosmological signal, which will in general be redshift dependent.   
In particular, we note that the dipolar and quadrupolar effects in (\ref{zdfirst2}) caused by the 4-acceleration of the observers in the expanding cosmological congruence description through the appearance of $\bm a$ in (\ref{def:h}) and (\ref{Picoefkappa}) are \emph{not} identical to the signatures of local aberration effects. 

We expect the dipole in the redshift drift signal to be dominated by the local secular aberration drift, while a quadrupole in data might be a signature of cosmological Weyl curvature/shearing effects, cf. (\ref{Pimultikappaapprox}) and the discussions below this equation. 
Local aberration effects might be corrected for separately, or alternatively be integrated in the cosmological analysis, by for instance employing a formalism for combining the full hierarchy of scales relevant for cosmological observation \cite{Korzynski:2017nas,Korzynski:2019oal}. 

{In order to apply cosmographic expressions of observational signals, the convergence and level of approximation of the Taylor series must be examined. 
In FLRW models, the radius of convergence of cosmographies is typically given by $\abs{z}=1$ due to a pole at the future null cone at $z=-1$ \cite{Lobo:2020hcz}. We might expect other expanding universe models to exhibit the same divergence of cosmographic expressions beyond redshifts of unity, since $z=-1$ generically corresponds to the physical singularity where the energy function of the photon tends to zero\footnote{See \cite{Cattoen:2007sk} for a suggested reparametrization of the redshift function appropriate for formulating cosmographies that are convergent for arbitrary redshifts in expanding FLRW universe models.}. 

The ratio of the second-order and first-order coefficients in the FLRW redshift drift cosmography is \cite{Lobo:2020hcz} $(j_0 - q_0^2)  /(2 q_0 )$, where $j_0$ is the FLRW jerk parameter as evaluated at the present epoch. For $j_0 \sim 1$ and $q_0 \sim 0.5$, this ratio of coefficients is of order unity, and the fractional error term of the first-order Taylor series is roughly $1 \times z$. Thus, a precision of $1\%$ (10\%) in the redshift drift cosmography can be achieved in an FLRW universe with $j_0 \sim 1$ and $q_0 \sim 0.5$ by considering redshifts no higher than $z\sim 0.01$ ($z\sim 0.1$). 

The convergence properties and level of approximation of the general first-order cosmography (\ref{zdfirst2}) must in principle be examined for each universe model of interest. The goodness of approximation of the truncated cosmography to the physical signal is in general expected to break down sooner for universe models where more structure on small scales is taken into account, giving rise to highly oscillatory signals. For a discussion of the convergence properties of the luminosity distance cosmography as a function of smoothing scale in the context of realistic general-relativistic universe simulations, see Appendix~A of \cite{Macpherson:2021gbh}. 

The higher-order polynomial terms in the cosmography (\ref{zdfirst2}) are naturally of interest. 
A challenge of formulating these in a representation useful for data analysis, is that gradients of the position drift, $\bm \kappa$, appear in the higher-order coefficients. 
The derivative of $\bm \kappa$ along the observer worldline is a second-order position drift effect and is therefore observationally challenging. Derivatives of $\bm \kappa$ would have to be dealt with for formulating useful higher-order expressions for redshift drift, either by reparametrising the expressions in terms of more accessible physical quantities or by making it plausible that higher-order position drift effects are subdominant in classes of realistic universe models. 
}

\section{Conclusion} 
\label{sec:conclusion} 
We have presented a framework for model-independent analysis of redshift drift data from nearby sources. 
Due to the representation of the redshift drift signal in terms of truncated multipole series in the incoming direction of the null ray and the position drift of the source as seen by the observer, the \emph{exact} anisotropic expression for redshift drift in the $\mathcal{O}(z)$ vicinity of the observer is given by a \emph{finite} number of coefficients. 

A fully model-independent analysis, making no use of complementary constraints or assumptions about the space-time congruence or the metric tensor of the Universe, implies 21 independent degrees of freedom to be determined. 
These degrees of freedom describe combinations of kinematic and curvature variables associated with the observer congruence and projections of the Ricci and Weyl curvatures of the space-time. 

The formalism allows {for transparently making} assumptions on the observer congruence description and the curvature of space-time, which might reduce the number of independent degrees of freedom to be determined from data. 
In the approximation of subdominant position drift and 4-acceleration of the congruence description, the number of independent degrees of freedom reduces to 11 -- and further reduces to six independent degrees of freedom in the case where the effective Hubble constant $\Eu_\obs$ can be measured by complementary data. 

In a general-relativistic hydrodynamic setting with subdominant anisotropic stress, these six independent degrees of freedom are given by the focusing term $u^\mu u^\nu R_{\mu \nu}$ (one degree of freedom) and the electric part of the Weyl tensor $u^\rho u^\sigma  C_{\rho \mu \sigma \nu}$ (five degrees of freedom). The number of degrees of freedom might be even further reduced; see the discussion below equation~(\ref{def:happrox}).  
The curvature invariants $u^\mu u^\nu R_{\mu \nu}$ and $u^\rho u^\sigma  C_{\rho \mu \sigma \nu}$ can in this case be determined directly from redshift drift data given sufficient constraining power for sources at low redshift. 

For an approximately isotropic sky sampling, the monopole term (given by the source term $u^\mu u^\nu R_{\mu \nu}$) of the redshift drift signal dominates, constituting a single scalar degree of freedom to be determined from data. 
The framework in the present paper has formal similarities with that presented in \cite{Heinesen:2020bej} for the analysis of distance-redshift data. The frameworks can be combined for joint model-independent constraints on the kinematics of the Universe.

\vspace{6pt} 
\begin{acknowledgments}
I wish to thank Chris Clarkson and Obinna Umeh for correspondence, and for making me aware that the results on the multipole decomposition of the effective deceleration parameter derived in \cite{Heinesen:2020bej} had previously been derived in Chapter 5 of the PhD thesis of Obinna Umeh \cite{Umeh:2013UCT}. I wish to thank Miko\l{}aj~Korzy\'nski for valuable comments, which helped improve this paper. I would also like to thank Thomas Buchert for his reading and comments on the manuscript, and Roy Maartens and Carlos Martins for correspondence regarding forecasts of redshift drift detections. 
{I thank Syksy~R\"as\"anen for correspondence on the interpretation of his published results on cosmic parallax, and Andrzej Krasinski for making me aware of the GRG published translation of the original 1927 paper by G.~{Lema\^itre} \cite{Lemaitre:1927}.}  
This work is part of a project that has received funding from the European Research Council (ERC) under the European Union's Horizon 2020 research and innovation programme (grant agreement ERC advanced grant 740021--ARTHUS, PI: Thomas Buchert).  
\end{acknowledgments}


\end{document}